# Near surface donor-acceptor pairs in hydrogenated homoepitaxial diamond nanolayers

A. M. Romshin[1,*], A.P. Bolshakov[1], A.A. Zhivopistsev[1], A. V. Gritsienko[2], O.S. Kudryavtsev[1], P. A. Pivovarov[1], V. G. Ralchenko[1], I. I. Vlasov[1]

*1- Prokhorov General Physics Institute of the Russian Academy of Sciences, 119991 Moscow, Russia*
*2 - P. N. Lebedev Physical Institute of the Russian Academy of Sciences, 53 Leninskiy Pr., Moscow 119991, Russia*

[*] - corresponding author

## Abstract

Hydrogen-terminated diamond is known for its *p*-type surface conductivity, which arises from a near-surface hole accumulation layer induced by adsorbed acceptor species. Here, we demonstrate that these surface acceptors also form optically active donor-acceptor pairs (DAP) with substitutional nitrogen donors in diamond. The insertion of a nominally undoped CVD interlayer between a nitrogen-rich HPHT substrate and a hydrogen-terminated surface enables the precise tuning of the donor-acceptor separation with nanometer precision. Radiative DAP recombination appears as bright, spectrally narrow lines whose intensity, energy, and decay dynamics depend systematically on interlayer thickness. Individual lines show single-photon statistics, while ensembles exhibit strong polarization anisotropy reflecting the planar donor-acceptor geometry. These findings reveal an optical counterpart of hydrogen-induced surface transfer doping in diamond and establish a surface-defined, nanometer-tunable platform for engineering DAP-based quantum emitters.

## Introduction

Hydrogen termination of the surface of pristine diamond has been regarded for more than two decades as a key factor initiating the emergence of high *p*-type surface conductivity [1–5]. Such diamond has attracted considerable research interest as a promising platform for the development of various types of electronic devices [6], including field-effect transistors [7,8], high-voltage vacuum power switches [9], and chemical sensors [10–12].

According to the transfer doping model proposed by Maier et al. [1], free holes are generated in the near-surface region of H-terminated diamond, while a negatively charged acceptor layer is formed on its surface, consisting of atmospheric adsorbates, most plausibly an adsorbed water layer involving the $H_3O^+/H_2O/H_2$ redox couple. Ristein et al. [13] established that substitutional nitrogen impurities, which act as deep donors in diamond, efficiently recombine with surface acceptors in H-terminated diamond at distances of 0-12 nm from the surface. It is known that the interaction between donor nitrogen impurities and acceptor boron impurities in the diamond bulk leads to radiative recombination of N–B pairs under optical excitation [14]. However, to the best of our knowledge, the optical properties of interacting “bulk nitrogen – surface acceptor” pairs have not yet been investigated, despite the fact that this interaction was detected by electrical measurements almost 25 years ago. First-principles studies of wide-bandgap materials with surface-induced acceptor states have predicted that near-surface donor-acceptor pairs can naturally acquire dipole moments aligned perpendicular to the surface while preserving favorable optical properties [15].

Recently, in nitrogen-doped H-terminated nanodiamonds, we observed a palisade of exceptionally narrow and spectrally bright lines in the 500-800 nm range, with a substantial fraction of these lines corresponding to single-photon emitters [16]. It was suggested that these emitters are associated with donor-acceptor recombination between nitrogen impurities and surface acceptors.

A distinctive feature of DAP recombination is the strong dependence of its spectral and kinetic properties on the donor-acceptor separation. In the classical model, the energy of the emitted photon is determined by the difference between the donor and acceptor state energies and the Coulomb interaction term of the pair $E_{DAP}(R) = E_D - E_A + \frac{1}{4\pi\varepsilon_0\varepsilon}\frac{e^2}{R}$, so that, as the separation $R$ increases, the Coulomb contribution decreases and the emission energy shifts to longer wavelengths. Moreover, in the dipole approximation, the radiative relaxation rate of DAP recombination is determined by the square of the optical matrix element, which describes the overlap of the localized donor and acceptor wave functions $W_{rad}(R) \sim exp(-\frac{2R}{a_{eff}})$, where $a_{eff}$ is the characteristic spatial scale of the overlap. Accordingly, $W_{rad}$ decreases with increasing donor-acceptor separation $R$ [17,18].

A systematic study of such dependencies in nanoparticles is challenging because the pair geometry is determined statistically and can be controlled only indirectly through the concentrations of donors and acceptors. In the present work, the focus is shifted from nanodiamonds to a planar geometry, and the initiation of donor-acceptor radiative recombination is observed in nominally undoped thin CVD layers grown on the surface of heavily nitrogen-doped HPHT single crystals, using laser excitation. This architecture differs fundamentally from nanoparticles in that it enables control over the distance between bulk nitrogen donors and surface acceptors through the thickness of the intermediate donor-depleted layer, and allows investigation of the distance dependence of the optical properties of radiative recombination.

## Results & Discussion

The creation of donor-acceptor pairs was achieved through the growth of a thin, nominally nitrogen-undoped CVD layer on the surface of a heavily nitrogen-doped HPHT single crystal (see **Fig. 1a**; see Methods for sample details). The epitaxial diamond films were deposited in the ARDIS-300 MPCVD reactor (Optosystems Ltd.) operating at 2.45 GHz using a $CH_4/H_2$ gas mixture [19]. During the process of growth, a portion of the substrate surface was covered with a smaller diamond mask, thereby shielding the underlying region from hydrocarbon radicals and reactive species in the hydrogen plasma.

This configuration resulted in a monotonically decreasing thickness profile of the nominally nitrogen-undoped CVD overlayer near the mask edge (see Methods for details) due to lateral gradients in radical's flux and, possibly, the substrate temperature.

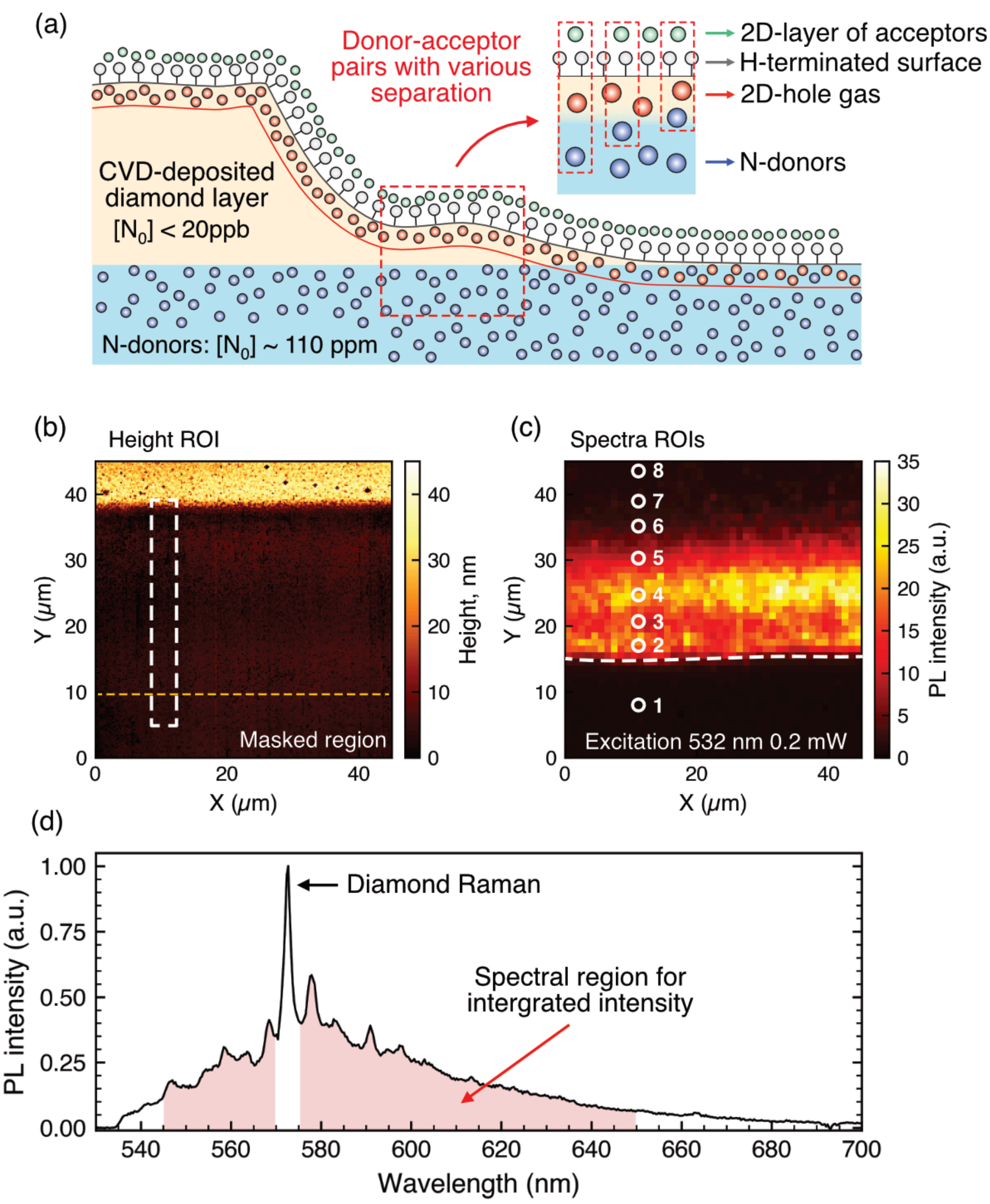

**Fig. 1**. (a) Schematic illustration of the formation of donor-acceptor pairs in nanometer-thick CVD layers that have been grown on a nitrogen-rich high-pressure high-temperature diamond substrate. (b) The topography of the formed nanoscale overlayer was measured by means of contact atomic force microscopy. The dashed rectangle marked Height ROI indicates the region from which the height profile shown in panel (e) was extracted. The bright elevated area in the upper part of the image corresponds to the masked region without CVD overgrowth, while the thickness-gradient overlayer is formed below it. (c) PL intensity map of the same area as on panel (b), integrated over the 545-650 nm spectral range and recorded under 532 nm excitation (500 μW). The circles labeled 1-8 mark the positions at which the representative spectra shown in **Fig. 3a** were recorded. (d) Representative PL spectrum measured at location 4 in panel (c), illustrating the emission of DAPs ensemble and spectral range that was used for integration of intensity.

Following the CVD overgrowth process, the sample underwent an additional treatment in hydrogen plasma for a duration of 10 minutes at a temperature of 750 °C to form a uniform hydrogen-terminated surface [20], both within the overgrown region and in the area that had previously been covered by the mask. Consequently, the distance between donor nitrogen in the bulk of the HPHT substrate and the surface acceptor states induced by hydrogen termination was controlled by the variable thickness of the intermediate CVD layer.

The structure was characterized using a combined technique integrating atomic force microscopy with confocal photoluminescence (PL) microspectroscopy, which enabled direct *in situ* correlation of the surface topography with the spatial distribution of PL in direction from masked substrate to CVD-deposited layer (see Methods). AFM measurements (**Fig. 1b, Fig. 2a**) demonstrate that the resulting profile is monotonic and comprises two characteristic regions: an extended shallow-slope region, with a height variation of approximately 6 nm over a length of ~25 μm, and a steeper region, with a height variation of approximately 30 nm over a length of ~3 μm.

Confocal PL mapping under 532 nm excitation revealed bright narrow-band emission in the spectral range of 545-650 nm (except the region of diamond Raman line 570-575 nm) across the entire profiled region (**Fig. 1c**), similar to that previously observed in nanodiamonds [16] and natural bulk crystals [21]. Concurrently, both the probability of detecting individual narrow-band emitters and their emission intensity exhibited a pronounced dependence on the local thickness $d$ of the overgrown layer. The most intense emission under 532 nm excitation was observed for layer thicknesses between 1 and 4.5 nm, with a maximum near 1.7 nm (**Fig. 2a,b**). In this thickness range, the PL maps exhibited the highest density of luminescent spots. In the limit of large $d$, the likelihood of detecting bright, narrow lines decreased significantly, and the intensity of the luminescence decreased accordingly. A similar suppression of emission was also observed in the limit of very small $d$, including the region corresponding to direct hydrogen termination of the surface without an intermediate undoped layer (**Fig. 2b**). The non-monotonic dependence of the emission properties on layer thickness indicates the existence of an optimal range of donor-acceptor separations in which the formation of bright emitters is most probable.

This phenomenon can be attributed to the interplay between radiative and non-radiative processes within DAPs, particularly at intermediate distances. In the case of small thicknesses, the shortest pairs predominate, for which the nearest donor is more strongly bound to the near-surface acceptor. The low intensity at large thicknesses is attributed to the weakening of radiative recombination caused by a reduction in the overlap of wave functions. Conversely, at the limit of small thicknesses, short-range configurations and the associated non-radiative quenching become dominant.

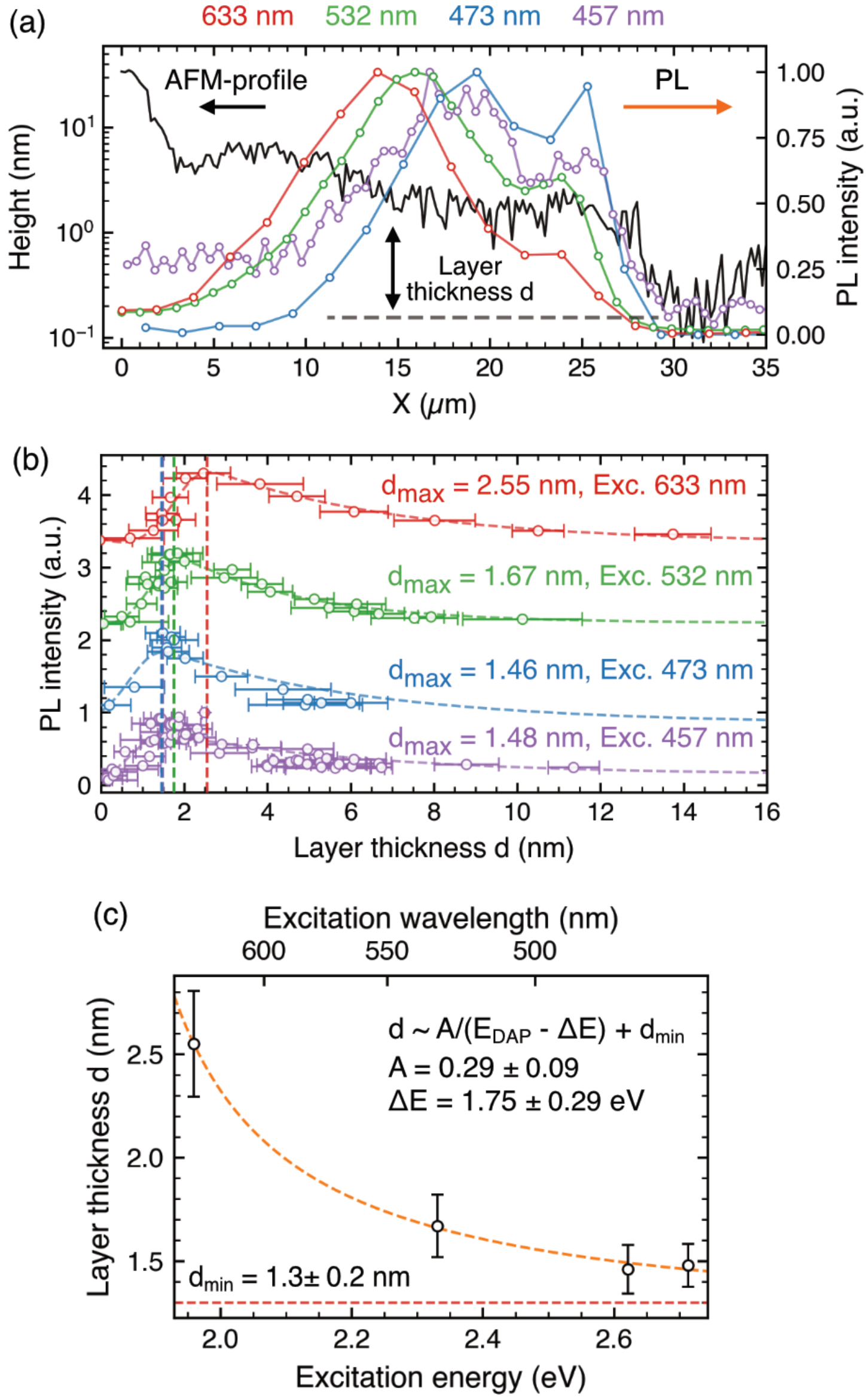

**Fig. 2**. PL intensity dependence on nanolayer thickness: (a) Height profile extracted from the region Height ROI marked in **Fig. 1b** (black curve), together with the corresponding PL intensity profiles recorded under excitations at 633 nm (red), 532 nm (green), 473 nm (blue), and 457 nm (purple) (see **Fig. S1** for maps recorded at 457, 473, 633 nm exc.). (b) Dependences of the DAP PL intensity on the overlayer thickness for different excitation wavelengths; the color notation is the same as in panel (a). Note the maxima in the thickness range of 1.5-2.5 nm. (c) Dependence $d_{I_{max}}(E_{exc} \approx E_{DAP})$ of the donor-acceptor separation on the excitation energy.

The observed intensity dependence on layer thickness can be naturally described within the framework of the hydrogen-like model of donor-acceptor recombination. In this model, $d$ is a parameter that defines the characteristic distance between a strongly bound bulk nitrogen donor and a weakly bound surface acceptor state [17,18]. In the present experiment, the excitation is of a selectively resonant nature, thus determining the recorded luminescence intensity as the product of the quantum yield of DAP recombination and the energy matching function between the excitation energy $E_{exc}$ and the pair energy $E_{DAP}(d)$: $I(d) \sim \eta(d) \cdot \zeta(E_{exc} - E_{DAP}(d))$, where $\eta(d) = \frac{W_r(d)}{W_r(d) + W_{nr}(d)}$ is quantum yield, $W_r(d)$ and $W_{nr}(d)$ are the radiative and non-radiative relaxation rates, respectively. The radiative contribution in the dipole approximation is assumed to decay exponentially with distance, $W_r(d) = W_0^r \cdot exp(-\frac{2d}{a_{eff}})$, which reflects the decrease in the overlap of the donor and acceptor wave functions and corresponds to the standard description of DAP recombination in the weakly-bound acceptor model [17]. By contrast, the non-radiative channel is treated as short-ranged and, for fitting the experimental data, is represented by a rapidly decaying exponential, $W_{nr}(d) = W_0^{nr} \cdot exp(-(\frac{d}{b})^2)$, where $b$ is the characteristic spatial scale of quenching. The matching function defines the range of DAP energies that are efficiently excited by the fixed-wavelength laser and, in the Gaussian approximation, can be written as $\zeta(E_{exc} - E_{DAP}(d)) = exp(-\frac{(E_{exc} - E_{DAP}(d))^2}{2\sigma^2})$, where σ is the energy width of the excitation window.

Similar measurements were performed under excitation at wavelengths of 457, 473, and 633 nm (**Fig. 2a**; see **Fig. S1** for PL maps). In all cases, the dependence of the integrated intensity on the thickness of the overgrown layer had a similar shape (**Fig. 2b**). However, the position of the intensity maximum systematically shifted to larger thicknesses with increasing excitation wavelength, occurring at 1.48 nm for 457 nm, 1.46 nm for 473 nm, 1.67 nm for 532 nm, and 2.55 nm for 633 nm. We attribute this behavior to selective resonant excitation of donor-acceptor pairs with different donor-acceptor separations as the excitation photon energy is varied: shorter-wavelength excitation addresses pairs with a larger Coulomb contribution and, consequently, a smaller donor-acceptor separation. In this case, the detected emission is not the resonant line itself but its phonon replica, shifted by approximately 180 meV (**Fig. 3a**). The dependence of the layer thickness $d$, corresponding to the maximum intensity, on the excitation energy $E_{exc} \approx E_{DAP}$, shown on **Fig. 2c**, can be expressed as a hyperbolic function $d_{I_{max}}(E_{DAP}) = \frac{A}{(E_{DAP} - \Delta E)} + d_{min}$ with $\Delta E = E_D - E_A$. From the obtained value $\Delta E \approx 1.75$ eV one can estimate $E_A$, taking into account that for positively charged nitrogen in the diamond lattice $E_D = 2.2$ eV [22]. This calculation provides a value of 1.45 eV for $E_A$ that corresponds to the magnitude of the valence-band bending near the diamond surface [1] and is in good agreement with estimates of such band bending reported in previous studies [23,24]. The non-zero value $d_{min} \approx 1.3$ nm reflects the deviation of $d(E_{DAP})$ from a purely hyperbolic form in the narrow near-surface region, where, owing to band bending, $\Delta E$ is no longer constant but becomes a function of $d$.

The spectra recorded at varying locations along the profile demonstrate a systematic evolution depending on the local layer thickness (**Fig. 3a**). In regions exhibiting thicknesses ranging from 1 to 3 nm, the maximum spectral density of emitters is observed. Individual narrow lines are situated in such close proximity to one another that they coalesce to form a broad, quasi-continuous ensemble, with an envelope maximum centered within the spectral range of 570 to 590 nm. The detuning of this maximum from the excitation frequency is 170-190 meV, which is in agreement with the values reported for DAPs in natural diamonds [21]. The broad band observed in the present work does not correspond to direct emission from resonantly excited DAPs, but rather to an ensemble of their phonon replicas. Its bell-shaped profile can be explained by the fact that, at a fixed excitation energy, not a single pair but a broad ensemble of pairs with similar transition energies is brought into resonance. Each of these pairs has its own phonon replica, and their cumulative superposition gives rise to the observed quasi-continuous profile.

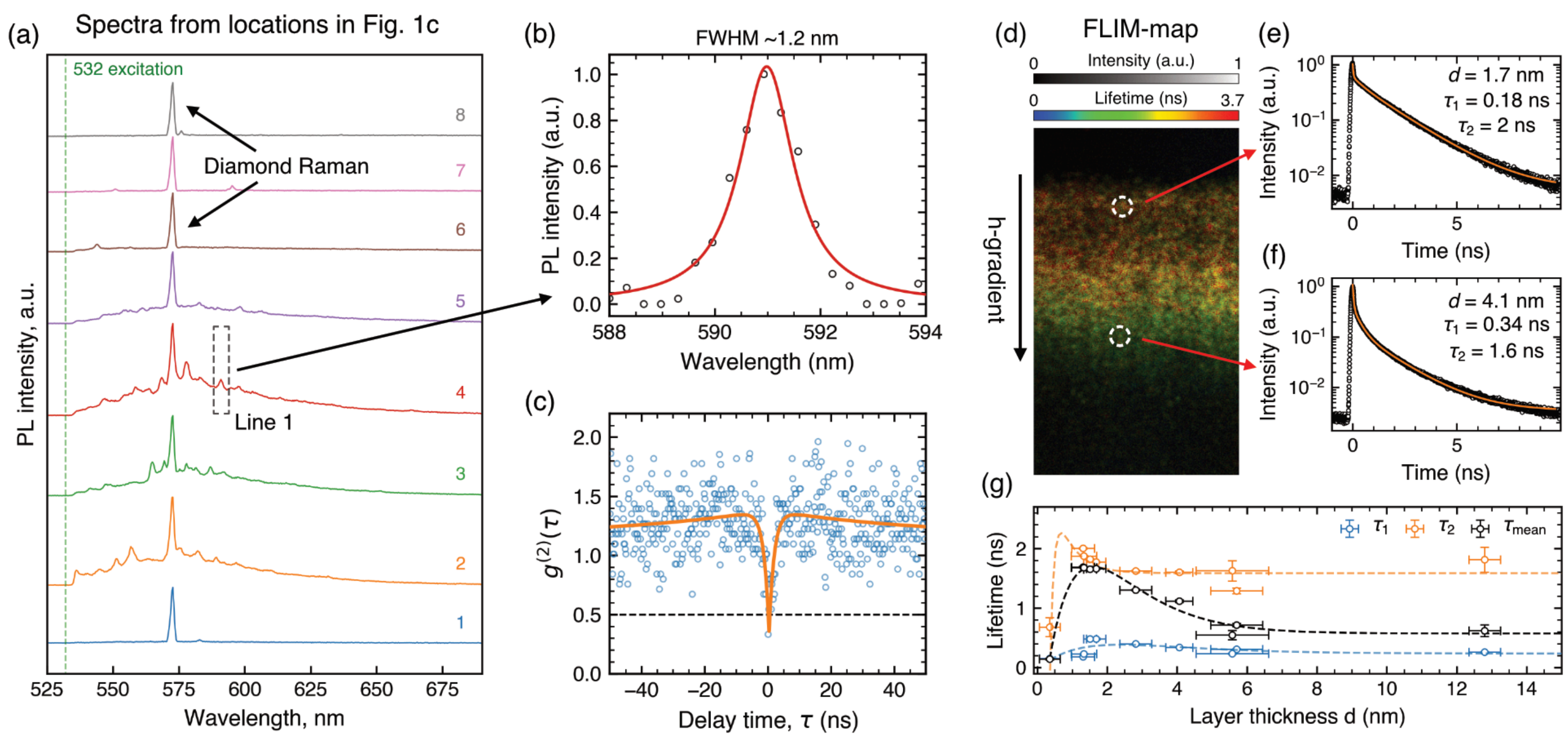

**Fig. 3**. (a) PL spectra recorded at eighth positions 1-8 labeled in **Fig. 1c**, corresponding to different overlayer thicknesses $d$, under 532 nm excitation (500 μW). (b) PL spectrum of a representative individual narrow line and (c) the corresponding second-order autocorrelation function, showing $g^{(2)}(0) = 0.37$. (d) Fluorescence lifetime imaging map (FLIM) of the selected ROI measured under pulsed 530 nm excitation in the 575-585 nm spectral window. (e, f) PL decay curves for two emitters formed in regions with different overlayer thicknesses, 1.6 and 4.1 nm, respectively. Solid lines show biexponential

fits. (g) Dependence of the extracted lifetimes $\tau_1$, $\tau_2$, and $\tau_{mean}$ on the overlayer thickness. The dashed black line shows an approximation of experimental data with the relation $\tau_{mean}^{-1} = W_r + W_{nr}$, where $W_r$ and $W_{nr}$ are the radiative and non-radiative rates introduced in the main text. The dashed blue and orange lines are splines employed for eye guidance.

PL lifetime measurements were performed under pulsed excitation with a 530 nm laser (see Methods) in order to investigate the recombination dynamics (**Fig. 3d-g**). The analysis concentrated on emission lines within the spectral range of 575-585 nm, corresponding to the region of strongest emission from the centers under study. For small overgrown-layer thicknesses of < 1 nm, the lifetimes were found to be shorter, on the order of 1.0-1.3 ns (see Methods). In this thickness range, the decay curves were described by biexponential functions. Increasing the layer thickness from 1 to 2 nm resulted in an increase in the average lifetime, reaching a maximum of 1.9 ns at a layer thickness of 1.7 nm, where the decay was closest to monoexponential (**Fig. 3e**). This thickness corresponds to the region in which the PL intensity reaches its maximum. A further increase in layer thickness resulted in a decrease in the average lifetime, which reached approximately 0.6 ns at larger thicknesses (> 6 nm).

This non-monotonic evolution of the lifetimes follows a similar thickness dependence as that observed for the PL intensity (**Fig. 2b**) and is consistent with the donor-acceptor recombination model, in which the thickness of the overgrown layer defines the characteristic separation between the bulk nitrogen donor and the acceptor state in the near-surface region. The total relaxation rate is governed by competition between radiative and non-radiative channels $\tau_{mean}^{-1} = W_r + W_{nr}$. In the limit of small $d$, the short lifetimes are associated with the dominance of short-range DAP configurations and the corresponding quenching, whereas near $d \approx$ 1.7 nm, radiative recombination is most efficient. With a further increase in layer thickness, the reduced overlap of the donor and acceptor wave functions leads to a decrease in $W_r$. Together with detuning from the resonant excitation condition, this decrease is accompanied by a reduction in the PL intensity and a shortening of the observed lifetime.

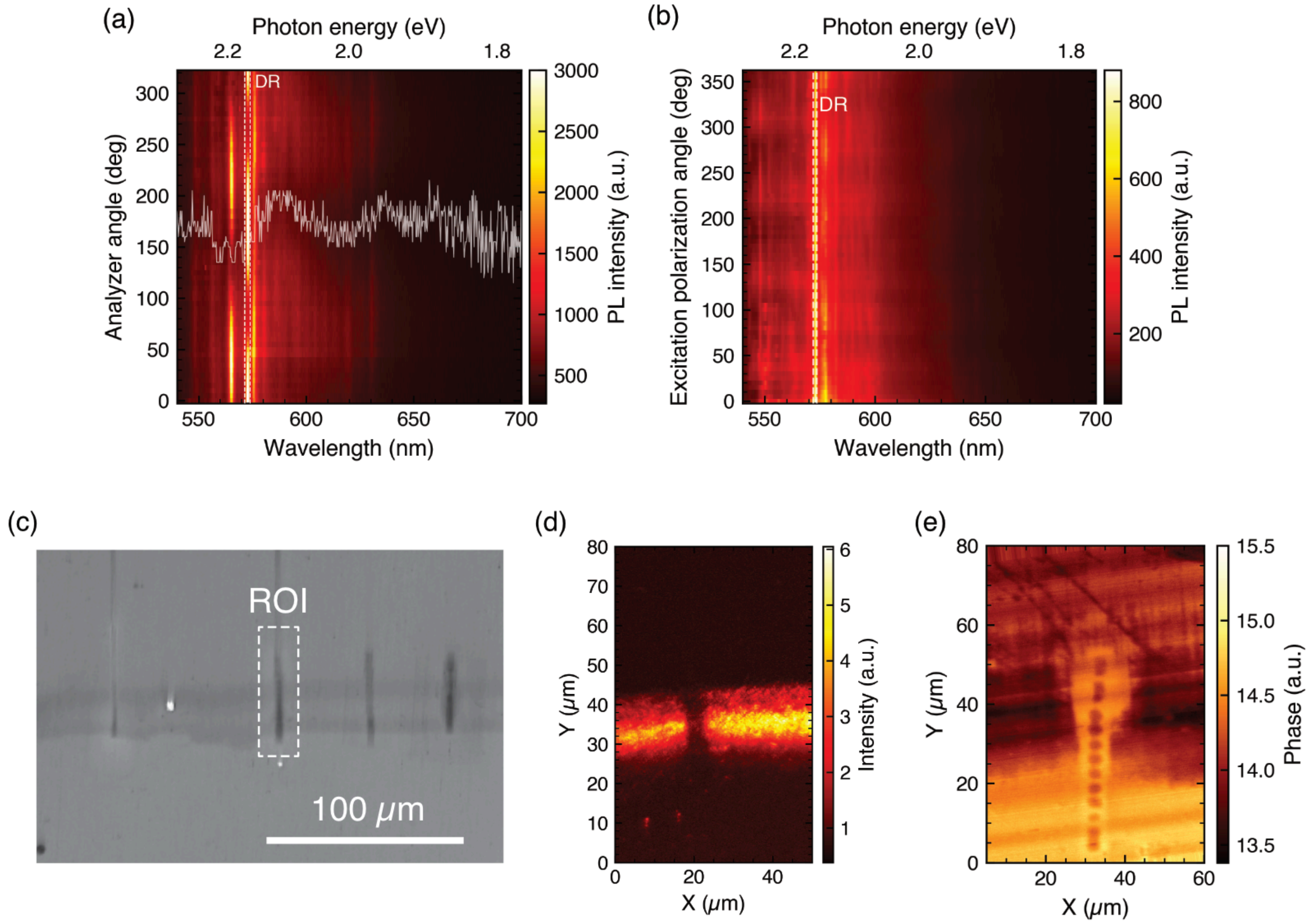

**Fig. 4**. (a,b) Polarization-resolved PL spectral maps of the DAP ensemble recorded under 532 nm excitation at 500 µW: (a) detection-polarization dependence and (b) excitation-polarization dependence. In panel (a), the white horizontal line marks the common analyzer angle at which the PL intensity of different narrow lines reaches its minimum, indicating a common orientation of the emitting dipoles within the ensemble. (c-e) Effect of electron-beam-induced chemical desorption of hydrogen from the diamond-film surface on the PL properties of DAPs: (c) enlarged SEM image of the regions exposed to the electron beam; (d) PL map of the region indicated in panel (c), recorded under 532 nm excitation at 1 mW using a 545-650 nm bandpass filter; (e) electrostatic force microscopy image of the same region.

For DAP ensembles exhibiting maximum intensity at $d = 1.7$ nm, polarization-resolved measurements were performed under 532 nm excitation (**Fig. 4a, b**). In the detection channel, rotation of the analyzer revealed a pronounced polarization contrast reaching 82%. By contrast, varying the polarization of the excitation light did not produce any noticeable changes in the PL intensity of the lines in the 545-700 nm range. Similar behavior was observed for two other DAP ensembles on the same sample.

This polarization behavior is typically observed for emitters whose dipole moments are oriented predominantly perpendicular to the surface. Given that the donor-acceptor pair is formed by a bulk donor and an acceptor state in the near-surface region, its effective dipole moment can be expected, on average, to be oriented along the surface normal. The close positions of the extrema in the polarization dependencies of the luminescence from different emitters further indicate that radiative transitions in DAPs with the same characteristic donor-acceptor separation have similar orientations, determined by the geometry of the pair with respect to the surface. The residual spread in orientations, not exceeding $\pm 20°$, is most likely associated with the detection of phonon replicas of resonantly excited DAPs, whose polarization response is modified by electron-phonon coupling and the resulting angular mismatch, as well as by the polarization sensitivity of the optical components.

The critical role of surface hydrogen termination in the observed emission was confirmed in a control experiment in which the sample was locally treated with an electron beam in a scanning electron microscope (**Fig. 4c**, see Methods for details), following the procedure described in [25,26]. After the treatment, the sample was re-examined by confocal PL microscopy. In the electron-beam-exposed region, the narrow-band emission was completely suppressed (**Fig. 4d**), demonstrating that local modification of the surface state eliminates the observed PL.

Electrostatic force microscopy mapping revealed an abrupt change in the phase contrast of the AFM-probe signal within the electron-beam-exposed region (**Fig. 4e**), indicating a significant modification of the near-surface electrostatic properties. Under an applied bias of +10 V between the probe and the sample, the treated region appeared as a dark area with a well-defined boundary. In the large-thickness region (>30 nm), the contrast was similar to that of the directly hydrogen-terminated surface without an intermediate layer, whereas the thickness range in which bright narrow-band emitters formed most efficiently exhibited a darker contrast. Bright halos, several micrometers to 5-8 μm wide, were also observed along the edges of the electron-beam-treated regions and are likely associated with local charge redistribution on the heterogeneously terminated surface.

Within the surface transfer doping model, the electrostatic response is governed by the finite density of surface acceptors responsible for band bending and the formation of a near-surface hole layer. In the thickness range corresponding to maximum DAP emission, a substantial fraction of these acceptor states may become involved in donor-acceptor pair formation, thereby reducing their contribution to the “free” hole reservoir. In this context, the darker contrast observed in the region of efficient DAP formation can be interpreted as evidence of partial compensation or binding of surface acceptors. This interpretation is qualitatively consistent with the results reported by Ristein et al. [13] and Kageura et al. [27].

The involvement of donor nitrogen in the observed PL was confirmed by control experiments using a pristine single-crystal diamond plate with a nitrogen concentration below 20 ppb as the starting sample. Under analogous conditions, a thin CVD layer was grown on this plate, followed by hydrogen-plasma treatment of the surface. In the resulting structure, PL similar to that observed for the heavily nitrogen-doped HPHT substrate was not detected (see **Fig. S2**). This result demonstrates that neither hydrogen termination alone nor the mere formation of a thin CVD layer is sufficient to produce the narrow-band emission considered here. Its appearance requires the presence of a nitrogen-doped substrate.

## Conclusion

In conclusion, we demonstrate for the first time that thin, weakly doped CVD overlayers grown on heavily nitrogen-doped HPHT diamond crystals provide a controllable platform for investigation of donor-acceptor recombination emission. In the design, characteristic separation between bulk nitrogen donors and hydrogen-induced near-surface acceptor states is defined by the thickness of the intermediate layer. We show that the emission intensity, spectral position, decay dynamics, polarization properties, and probability of observing individual single-photon emitters all exhibit a pronounced non-monotonic dependence on this parameter. Control experiments based on local hydrogen desorption and low-nitrogen reference samples unambiguously identify substitutional nitrogen and hydrogen-induced surface acceptors as the two essential components of the observed emission. Importantly, the brightest DAP ensembles exhibit a high degree of orientational uniformity in their emissive transitions, imposed by the geometry of the bulk-surface pairs. Taken together, these results establish the DAP origin of narrowband emission in hydrogen-terminated diamond and open a route toward ordered arrays of identically oriented quantum emitters compatible with diamond nanophotonics and scalable solid-state quantum-network architectures.

## Acknowledgements

This study was supported by the Ministry of Science and Higher Education of the Russian Federation (Agreement/Grant No. 075-15-2025-609).

## Materials and methods

### 1. *Single crystal diamond substrates*

Two diamond samples were used in this work. The main sample was a strongly yellow HPHT single-crystal diamond plate with a (100)-oriented surface (sample 1). As a reference sample, we used a single-crystal CVD diamond plate supplied by Element Six, with the same surface orientation and a substitutional nitrogen concentration below 20 ppb (sample 2).

The nitrogen concentration in the HPHT sample was determined by Fourier-transform infrared (FTIR) spectroscopy (PerkinElmer Spectrum 100) from the intensity of the absorption bands associated with nitrogen-related defects. First, the absorption spectrum was recorded over the 1000-4000 $cm^{-1}$. range. The spectrum was then normalized so that the intrinsic two-phonon absorption band of diamond had an intensity of 14. This normalization converts the measured spectrum to the equivalent spectrum of a 1-mm-thick sample, since the calibration coefficients relating absorption intensity to defect concentration were established for diamond plates of that thickness.

The concentration of single substitutional nitrogen $N_s$ (C centers) was estimated from the absorption band at 1135 $cm^{-1}$ using the standard relation $n_c = 4.4 \cdot 10^{18} \cdot \mu_{1135}$ $cm^{-3}$, where $\mu_{1135}$ is the normalized absorption intensity at 1135 $cm^{-1}$. For the measured value $\mu_{1135} = 4.62$ this gives $n_c = 2.0 \cdot 10^{19}$ $cm^{-3}$, which corresponds to approximately 110 ppm of substitutional nitrogen.

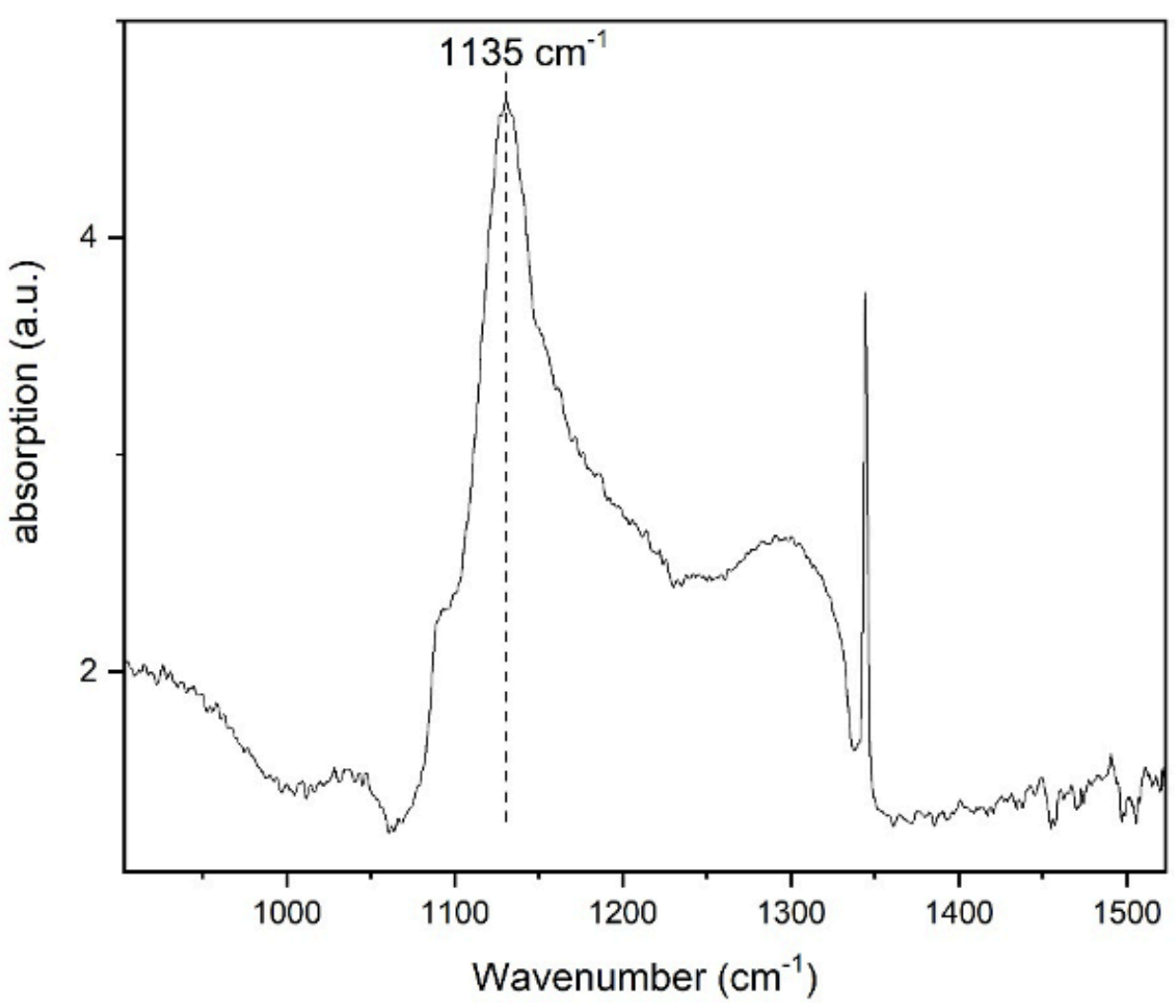

**Fig. A1**. IR-absorbance spectrum of C-centers in HPHT diamond substrate.

*2. Homoexpitaxial growth of hydrogenated diamond nanolayer*

Thin epitaxial diamond films (2–5 nm) were synthesized in an ARDIS-300 MPCVD reactor (Optosystems Ltd.) operating at 2.45 GHz (up to 6 kW) in a $CH_4/H_2$ gas mixture. Purity of the process gases: $H_2$ – 99.99999 vol.% and $CH_4$ – 99.9999 vol.%. Growth was carried out on HPHT single-crystal substrates with a (100) working surface orientation. Prior to the growth session, the substrates were subjected to a standard cleaning procedure in acetone in an ultrasonic bath, and then placed inside the reactor's vacuum chamber on a molybdenum substrate holder. To create a thickness profile for the synthesized CVD film, a diamond mask – a small-area (~1×3 mm) single-crystal polished plate – was placed on the growth surface, covering part of the substrate surface. The synthesis conditions varied within the following range: chamber pressure 60 – 70 Torr, microwave power 2 – 2.3 kW, methane content in the methane-hydrogen mixture 0.5%, substrate temperature 800 – 830

°C. The substrate temperature was measured with a two-colour pyrometer (METIS M322 model, SensorTherm) with spatial resolution of 0.8 mm. A distinctive feature of growing such thin films is the short synthesis time: 30 – 8 seconds. In this case, the microwave plasma must be switched off abruptly without reducing the power or pressure in the chamber, as is usually done during standard CVD synthesis of films tens or hundreds of micrometers thick.

After CVD synthesis, the diamond mask was removed and the sample (the initial substrate with the grown CVD film) was treated in a hydrogen plasma in the same reactor for 10 minutes under growth conditions, but without the addition of methane, to create a quasi two-dimensional layer of acceptors.

### 3. *Atomic force and confocal PL microscopy measurements.*

AFM measurements were performed using a commercial AFM microscope integrated with a confocal optical module (NTEGRA Spectra, NT-MDT). A contact-mode AFM probe was first used to record the surface height map of the sample. After completion of the topography measurements, the probe was retracted and spatially resolved PL spectral maps were recorded from the same region using excitation wavelengths of 457, 473, 532, and 633 nm. The emitted signal was analyzed with a monochromator (Solar M550) equipped with a CCD camera (Andor Newton).

Single-photon measurements of individual DAP emitters were carried out using a custom-built optical microscope. Precise positioning of the sample with respect to the laser focus was achieved with a three-axis piezoelectric stage (piezosystem jena) monitored within the field of view of a CMOS camera. Excitation was provided by a 532 nm laser with a power of 0.5 mW at the sample. Both excitation and collection were performed with a Mitutoyo 100x objective (*NA*=0.7). The collected PL was filtered by a combination of notch, angle-tuned long-pass, and angle-tuned short-pass filters (Semrock) and directed to the detection path. For PL characterization, avalanche photodiodes (APDs, Excelitas SPCM-AQRH-14-FC) were used in combination with notch and band-pass filters. Individual narrow lines were spectrally selected using angle-tuned filters. The single-photon character of the DAP emission was verified with a Hanbury Brown Twiss interferometer employing two APDs of the same type.

Time-resolved measurements were performed using a second optical setup based on a PicoQuant MicroTime 200 fluorescence confocal microscope coupled to an Olympus IX71 inverted microscope. The emitters were excited by 532 nm pulses from a supercontinuum laser (SpectraK, NKT Photonics) with an average power of approximately 60 μW, repetition rate of 80 MHz, and pulse duration of 50 ps. The emitted signal was detected with a single-photon avalanche photodiode (Micro Photon Devices, PDM series, 35 ps timing resolution) and processed with a PicoHarp 300 counting module (PicoQuant). A band-pass filter in the 575-585 nm range was used to isolate the spectral region of interest and suppress the remaining luminescence background.

*4. Decay time estimations*

The experimental fluorescence decays demonstrate a multiexponential behavior and can be approximated by a sum of several exponential components [1]:

$$I(t) = \sum_i a_i e^{-t/\tau_i},$$

where $a_i$ and $\tau_i$ are the intensity-weighted amplitude and fluorescence decay time for the $i$-th component. To compare the decay times for the DAP emitters in different overlayer thicknesses, we use the quantity:

$$\tau_{mean} = \frac{\sum_i a_i \tau_i^2}{\sum_i a_i \tau_i},$$

where $\tau_{mean}$ is the average fluorescence decay time. The approximation of the experimental curves by the sum of two exponential components allows us to determine the average fluorescence decay time. We take into account the instrument response function (IRF), with its full width at half maximum being ~90 ps at 580 nm, as done in [29]:

$$I_{exp}(t) = \int_{-\infty}^{\infty} IRF(t')I(t - t')dt',$$

where $I_{exp}(t)$ is the decay observed in the experiment.

*5. Hydrogen desorption in SEM.*

Local modification of the hydrogen-terminated surface was carried out in a scanning electron microscope by electron-beam irradiation at an accelerating voltage of 15 kV, beam current of 1 nA, and exposure time of 30 s. Square rasters of 3x3 $\mu m^2$ were written sequentially with a translational step of 100 μm along the thickness gradient of the overgrown layer in order to probe regions with different donor-acceptor separations.

## Supplementary Information

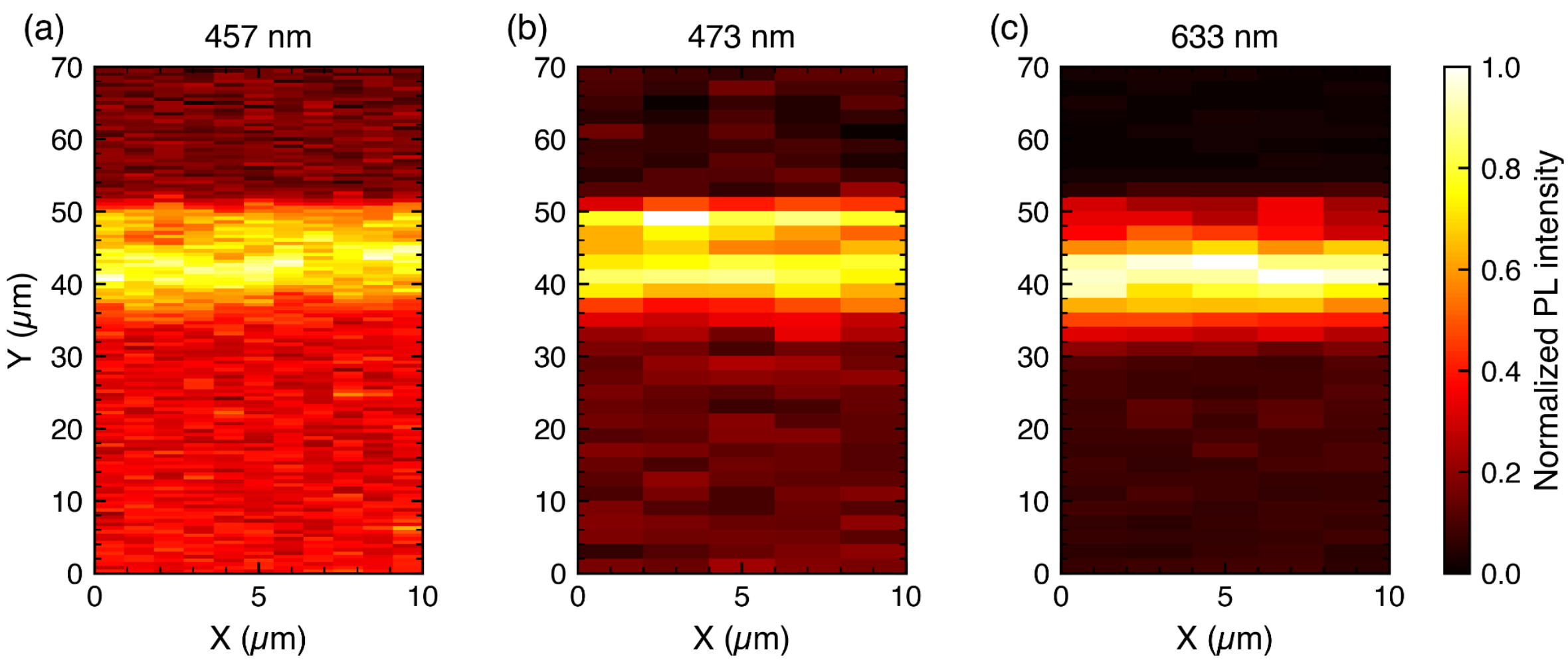

**Fig. S1**. Photoluminescence maps of the same region of interest as in Fig. 1b, recorded under (a) 633 nm, (b) 473 nm and (c) 457 nm excitations.

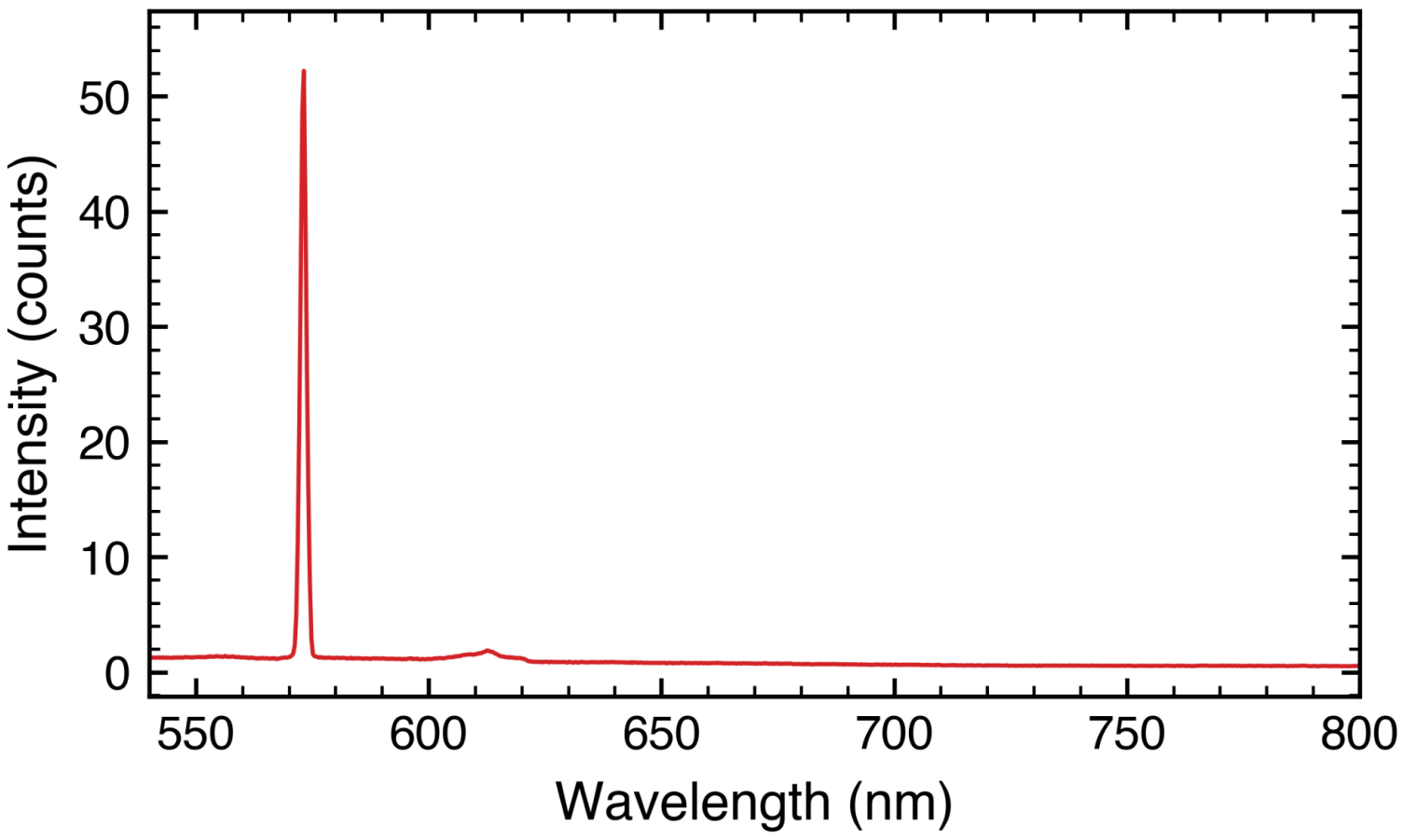

**Fig. S3**. Photoluminescence spectra of the hydrogenated CVD nanolayer grown on a single-crystal diamond substrate with a substitutional nitrogen concentration below 20 ppb with no sign of DAP radiative recombination. Recording was conducted under the same conditions as for DAPs in sample 1. The narrow line at 572.6 nm corresponds to first-order diamond Raman scattering, an imprint at the 600-620 nm corresponds to second-order diamond Raman.